\documentstyle[prb,aps,graphics]{revtex}

\newcommand{\FIG}[1]{#1}
\newcommand{\PREP}[1]{}

\newcommand{\vv}{\mbox{\bf v}}
\newcommand{\BB}{\mbox{\bf B}}

\begin{document}
\draft
\title{Non-linear dynamics of Kelvin-Helmholtz unstable magnetized jets:
three-dimensional effects}
\author{R. Keppens}
\address{FOM-Institute for Plasma Physics, 
        P.O.~Box 1207, 3430 BE Nieuwegein, The Netherlands} 
\author{G. T\'oth}
\address{Department of Atomic Physics, E\"{o}tv\"{o}s University, 
         P\'azm\'any P\'eter s\'et\'any 1, Budapest H-1117, Hungary}
\maketitle

\begin{abstract}
A numerical study of the Kelvin-Helmholtz instability 
in compressible magnetohydrodynamics is presented. 
The three-dimensional 
simulations consider shear flow in a cylindrical jet configuration,
embedded in a uniform magnetic field directed along the jet axis.
The growth of linear perturbations at specified poloidal and
axial mode numbers demonstrate intricate non-linear coupling effects.
The physical mechanims leading to induced secondary 
Kelvin-Helmholtz 
instabilities at higher mode numbers are identified. The initially weak magnetic
field becomes locally dominant in the non-linear dynamics before and
during saturation. Thereby, it controls the jet deformation and eventual 
breakup. 
The results are obtained using the Versatile Advection Code
[G.~T\'oth, Astrophys. Lett. \& Comm. {\bf 34}, 245 (1996)], 
a software package designed to solve general systems of conservation laws. 
An independent calculation of the same Kelvin-Helmholtz unstable
jet configuration using a three-dimensional 
pseudo-spectral code gives important insights into 
the coupling and excitation events of the various linear mode numbers.
\end{abstract}


\section{Introduction}\label{s-intro}

The Kelvin-Helmholtz (KH) instability is a basic hydrodynamic phenomenon
in sheared flows~\cite{chandra,blumen}. Magnetic fields can strongly
influence the linear and non-linear behavior of this instability. A recent
full parameter study~\cite{vac-kh2d} (Paper I)
of the growth and saturation of the KH instability in two-dimensional (2D)
compressible magnetohydrodynamics (MHD) augmented
the vast body of knowledge on magnetically induced 
effects~\cite{frank,jones,dahl97,dahl98,malag,miura,miupr,min}. 
It illustrated how a weak, uniform $\BB$
field gets amplified by the developing
vortical flow and how this in turn changes the
redistribution of mass as compared to pure hydrodynamic cases. 
Figure~\ref{f-2d}, taken from Paper I, shows a close-up of the
density pattern after four transverse sound travel times in a magnetically
modified KH evolution. Parameter values are listed below
in section~\ref{ss-iv}. The magnetic field becomes dynamically dominant
at the time shown, halting the further growth in the transverse $y$-direction.
The $x$-direction is periodic, so Fig.~\ref{f-2d} shows alternating
regions of density enhancements (bright) and depletions (dark). The narrow
lanes of low density intersecting the high density regions signal sites
of strong magnetic fields. 

The KH instability is important for numerous astrophysical applications.
Shear flows occur in jets ejected from 
star forming regions, in stellar winds, in winds emanating from accretion
disks, \ldots Many of the jet-type astrophysical flows 
are at such high speeds that the dominant dynamics
is through shock interaction both internal to the jet and at its leading
working surface where the jet penetrates the ambient plasma. The fundamental
shear-induced instabilities that can develop at the jet surface 
are ultimately responsible for entrainment and mixing with the ambient material.
Recently, Bodo {\it et al.}~\cite{bodo} performed comparative 2D and 3D hydrodynamic
simulations of supersonic (Mach 10) jets. In their 
3D simulations of a jet, 10 times less dense than its environment, 
faster decay into small-scale 
structure was demonstrated, as well as strong mixing with the external medium.
The evolution was governed by the formation of internal shocks, induced
external shocks, and consequent momentum and material mixing processes. 
Usually, the initial perturbation in 2D and 3D
studies of this kind consists of a superposition of all linear wave modes,
mimicking the dynamics resulting from random excitations. It is then very
difficult to disentangle what causes the separate modes to couple. Therefore,
we take a different approach and perturb at selected mode pairs. Our aim is
to clearly identify cause and effect in the non-linear dynamics. 

Typically, highly collimated jets are seen as bipolar
outflows from young stellar objects. It is generally
believed that the magnetic field plays an as yet not fully understood role
in this collimation. Insights from 2D simulations have been carried over
to 2.5 dimensions~\cite{jones,miura,frank2}. 
Numerical, fully three-dimensional calculations of
astrophysical phenomena involving shear flow and
magnetic fields consider MHD simulations of relativistic jets in 
aligned~\cite{nishi1} and oblique magnetic fields~\cite{nishikawa},
and have focused on density effects in 
poloidally magnetized supermagnetosonic jets~\cite{hardee}.
In contrast to these studies where the long-term coherence of the jet
is of interest, we assess initial transient features and the way in which
Lorentz forces can saturate and control the jet deformation and breakup.
We highlight this role played by magnetic
fields in a simple cylindrical jet configuration with a
uniform field parallel to the shear flow. Our study of 3D jets looks at these
magnetic effects in the subsonic Kelvin-Helmholtz unstable regime. 
We concentrate on the three-dimensional non-linear hydrodynamic and magnetic
effects. Due to the large computational cost required for 3D simulations,
we restrict ourselves to two cases which are an immediate 3D generalization 
of the 2D case shown in Figure~\ref{f-2d}. 
The governing parameters~\cite{vac-kh2d} 
are chosen such that (i) the magnetic 
field is weak initially, allowing the hydrodynamic Kelvin-Helmholtz
instability to develop; and 
(ii) the shear flow strength corresponds to
its most unstable value in 2D in the sense 
that the saturation level is highest at fixed axial mode number.
This axial wave number is taken close to, but slightly below
the most unstable linear mode from the 2D case.
By setting the mode number in the azimuthal direction we can investigate
the influence of the three-dimensional geometry.

In this parameter regime, we can simulate the dynamics using 
two entirely different numerical tools: the general finite-volume
based Versatile Advection Code (VAC~\cite{tothcrete,tothvienna}), 
and the combined finite-difference, spectral code HEating by Resonant 
Absorption (HERA~\cite{hera-hpcn97,hera-hpcn98}). 
While the latter is designed
for magnetized loop dynamics involving strong radial gradients only, 
the former offers a choice of high-order, shock-capturing schemes 
typically used in computational fluid dynamics calculations. 
Of these two codes, only VAC is able to
investigate the supersonic regime relevant for astrophysical jets.
In this combined VAC-HERA study of the subsonic regime, we not only
verify both codes against eachother, but also 
get additional insight in the various spectral contributions to the dynamics.

The MHD equations, initial conditions, and the numerical tools used 
are discussed in section~\ref{s-eq}. We limit the discussion of the non-linear
dynamics to two cases that only differ in the perturbation applied to the 
KH unstable jet. Section~\ref{s-dynamics} contains a detailed analysis of
their saturation behavior and illustrates the role played by the magnetic 
field.
We briefly comment on the further evolution. Conclusions are presented in 
section~\ref{s-concl}.

\section{Numerical modeling of the problem}\label{s-eq} 

The ideal MHD equations constitute a set of conservation laws. The
eight partial differential equations govern the
time evolution of mass density $\rho$, momentum density $\rho \vv$, 
total energy density $e$, and magnetic induction $\BB$. 
Written in these conservative variables, we get
\begin{equation}
\frac{\partial \rho}{\partial t}+\nabla \cdot (\rho \vv)=0,
\end{equation}
\begin{equation}
\frac{\partial (\rho \vv)}{\partial t}+ \nabla \cdot [ \rho \vv \vv + p_{tot} I-
\BB \BB]=0,
\end{equation}
\begin{equation}
\frac{\partial e}{\partial t}+ \nabla \cdot (e \vv) + \nabla \cdot (p_{tot} \vv) - \nabla \cdot (\vv \cdot \BB \BB )= 0,
\label{q-e}
\end{equation}
\begin{equation}
\frac{\partial \BB}{\partial t}+
\nabla \cdot (\vv \BB-\BB \vv)= 0.
\label{q-b}
\end{equation}
We introduced 
$p_{tot}=p + \frac{1}{2}B^2$ as the total pressure,
and $I$ as the identity tensor.
The thermal pressure $p$ is related to the energy density as
$p=(\gamma-1)(e-\frac{1}{2}\rho v^{2}-\frac{1}{2}B^{2})$.
We set the adiabatic gas constant $\gamma$ equal to $5/3$.
Magnetic units are defined such that the magnetic permeability is unity.

\subsection{Numerical tools}\label{ss-tools}

The numerical 3D time-dependent calculation 
of the Kelvin-Helmholtz instability in magnetized plasma jets is
done twice with two completely independent software tools, the Versatile
Advection Code (VAC~\cite{tothcrete,tothvienna}), 
and the HEating by Resonant Absorption code 
(HERA~\cite{hera-hpcn97,hera-hpcn98}). Detailed descriptions of the 
algorithms employed in these codes are found through the references, here we
only highlight the differences. VAC is more general 
in terms of algorithms~\cite{tothodstrcil,implvacA}, 
applications~\cite{implvacB,wind98,vac-kh2d,porto98}, 
and geometry, while HERA is truly 
specialized to the solution of the 3D MHD equations in cylindrical 
geometry~\cite{poedts,hera-hpcn98}.
Both codes run on vector and distributed memory architectures, and their
implementation and performance aspects are detailed 
elsewhere~\cite{hpfart,vac-hpcn98}.
The VAC and HERA results in this paper were obtained on a shared memory
vector Cray C90 and on 16 processors of the Cray T3E and IBM SP.

VAC employs a finite volume discretization on a structured grid, 
while HERA uses finite differences in the radial direction, and a 
2D spectral
\begin{equation}\label{q-spectral}
  \exp\left(i m \varphi +i n \frac{2\pi Z}{L}\right)
\end{equation} 
representation in the azimuthal direction ($\varphi$)
and along the cylinder ($Z$). In VAC, we selected the explicit, 
one-step Total Variation Diminishing scheme~\cite{harten}
with {\it Woodward} limiting~\cite{collwood}.
This scheme makes use of a Roe-type approximate Riemann solver~\cite{roe}.
The conservation form of the MHD equations is taken and
ensured by this discretization.
HERA uses a semi-implicit predictor-corrector
time stepping algorithm~\cite{hera-hpcn97} and advances the primitive
variables $\rho$, $\vv$, $p$, and $\BB$. 
The discrete equivalent of $\nabla \cdot \BB =0$ is
ensured in VAC by applying a projection scheme at every time 
step~\cite{brackbarn} using an efficient iterative method. 
HERA instead uses a staggered radial mesh,
which allows to encode a discretization that keeps the
field divergence identically zero. For the HERA calculations reported in 
section~\ref{s-dynamics}, we use a stabilizing `artificial viscosity' term 
$\mu\nabla^2 \vv$ added to the momentum equation, with $\mu={\cal O}(10^{-4})$.
For the resolutions used in the calculations, this turns out to be
roughly at the order of the numerical dissipation inherent to the 
scheme used from VAC.

\subsection{Initial conditions}\label{ss-iv}

The 3D jet configuration generalizes the 2D simulation shown in 
Figure~\ref{f-2d}. This calculation started from
a uniform density $\rho_0=1$, pressure $p_0=1$ and magnetic field $\BB_0$
parallel to a shear flow of the form $v_x=0.645\tanh(y/0.05)$. 
The initial plasma beta was $\beta=2 p_0/B^2_0\simeq 120$, and this
2D reference case is
further characterized by the dimensionless
sound Mach number $M_s=0.5$ (ratio of the shear flow amplitude to the
sound speed) and Alfv\'en Mach number $M_a=5.0$ (ratio of shear
flow amplitude to the Alfv\'en speed). Our previous study in two space
dimensions~\cite{vac-kh2d} showed that 
for these parameters, the non-linear saturation level is
maximal at the chosen sound Mach number for the mode 
with wave number $k=1$. Therefore we apply a perturbation to the shear flow
according to $\delta v_y \sim \sin (2\pi x)$.

In 3D, we consider a jet-like flow.
Using $(R,\varphi,Z)$ cylindrical coordinates, we set up
a flow field 
\begin{equation}
   v_Z(R; t=0) = V_0 \tanh \frac{R-R_{jet}}{a}
\end{equation}
parallel to a cylinder of length $L=1$ and radius $R_{jet}=0.5$,
which is sheared in the radial direction. Note that the jet surface $R=R_{jet}$
is easily identified as the $v_Z=0$ isosurface.
The strength of the velocity shear is fixed at $V_0 =0.645$,
and the width of the shear layer is $a = 0.1 R_{jet}=0.05$
exactly like in the reference 2D case.
Further, the initial pressure $p_0$ and density $\rho_0$ are equal to unity
everywhere, and we impose
a uniform initial field  $\BB=B_0 \hat{e}_Z$ parallel to the jet.
Taking $B_0=0.129$, we have a 3D configuration with identical properties
as the reference KH unstable 2D case~\cite{vac-kh2d} shown in Fig.~\ref{f-2d}. 
We perturb the cylindrical jet surface by imposing a radial velocity
profile 
\begin{equation}
  v_R(R,\varphi,Z; t=0)=
  \Delta v_R \exp{-\left(\frac{R-R_{jet}}{4a}\right)^2}
  \cos(m\varphi) \sin\frac{n2\pi Z}{L}. 
\end{equation}
Hence, we impose a perturbation at specific mode numbers $m$ and $n$.
This does represent a truly 3D perturbation,
as it translates into the simultaneous excitation of four 2D Fourier
modes (\ref{q-spectral}), namely $(+m,+n)$ with complex
amplitude $-i/4$, $(+m,-n)$ with amplitude $+i/4$, and similarly for
the $(-m,\pm n)$ modes (since the velocity is a purely real quantity). 
These symmetry considerations will allow us to refer to $(m,n)$ mode
pairs with always $m\geq 0$ and $n\geq 0$, tacitly assuming the
$(\pm m, \pm n)$ individual mode contributions.
The amplitude of the perturbation is fixed at $\Delta v_R = 0.01$ and 
its radial width is chosen to be $4a=0.2$.

\subsection{Boundary conditions}\label{ss-bc}

We calculate the evolution of the KH unstable jet with both VAC and HERA.
For VAC, we set 
up the jet configuration in a Cartesian $(x,y,z)$ box
of sizes $L\times 4 R_{jet} \times 4 R_{jet}$. 
Note the difference between the Cartesian $z$ and the cylindrical 
$Z$-direction.
In fact, the first, periodic $x$-dimension is now along the jet.
We assume outflow boundary conditions on the
sides parallel to the jet. We take $L=1$ and 
an aspect ratio of $L/R_{jet}=2$. We use $50 \times 100 \times 100$ equidistant
grid points.

For HERA, the jet is placed within a co-axial cylinder of radius $2R_{jet}$.
We took $64$ radial grid points, and 32 modes both
in the $\varphi$ and $Z$ dimensions.
HERA naturally assumes periodic boundary conditions along ($Z$-direction) 
and about ($\varphi$-direction)
the loop, while a regularity condition holds at $R=0$. We took a
closed, perfectly conducting wall at the outer radius, 
which is the only difference in the problem setup between VAC and HERA.
All calculations reported below are not (yet) influenced by this
difference in `open box' versus `closed cylinder' outer boundary configuration. 

\section{Non-linear dynamics in sheared jets}\label{s-dynamics}

\subsection{Case $m=1$}

Figure~\ref{f-vachera} shows the time evolution of the scaled
poloidal kinetic and magnetic energies from both calculations
when we perturb at mode numbers $m=1$ and $n=1$.
These are volume integrated quantities, namely
\begin{equation}
 E^{pol}_{mag}(t)\stackrel{\rm VAC}{=}
\frac{1}{B_0^2 R_{jet}^2 L}\int_V dV\,\frac{B_y^2+B_z^2}{2}
\stackrel{\rm HERA}{=}
\frac{1}{B_0^2 R_{jet}^2 L}\int_V dV\,\frac{B_R^2+B_{\varphi}^2}{2}, 
\end{equation}
and similarly
\begin{equation}
 E^{pol}_{kin}(t)\stackrel{\rm VAC}{=}
\frac{1}{\rho_0 V_0^2 R_{jet}^2 L}\int_V dV\,\rho\frac{v_y^2+v_z^2}{2}
\stackrel{\rm HERA}{=}
\frac{1}{\rho_0 V_0^2 R_{jet}^2 L}\int_V dV\,\rho\frac{v_R^2+v_{\varphi}^2}{2}, 
\end{equation}
and they allow a determination of the growth rate and the saturation
behavior of the instability. Note that the difference in the total volume
of the simulated box (VAC) or cylinder (HERA) does not make a difference
for these volume integrals since the functions are zero far from the jet
surface.

The agreement between the finite volume based VAC code and the
pseudo-spectral HERA code is excellent: the growth and saturation behavior
is clearly captured at about $t=4$. The spectral code allows us to 
identify the contributions of all linear wave numbers $(m,n)$ at any
particular moment in time. The contributions to the poloidal
magnetic energy of the most dominant wave numbers are indicated in the
figure. We immediately read off how the $(1,1)$ perturbations induce
$(0,2)$, $(2,2)$, and $(2,0)$ modes, in that order. These couple and
reverse their original ordering at $t\simeq 0.75$. At saturation $t=4$, 
the three most important mode number pairs are the excited
$(1,1)$ and the induced $(3,1)$ and $(0,2)$ modes.
Note that, interestingly, the $(m,m)$ modes with $m\geq 2$
of similar helicity as the excited $(1,1)$ mode never play an important
role during and beyond the saturation phase, at least not for $t\leq 6$. 
This is in spite of
the fact that they are excited fairly early on in the growth phase of the
instability: the $(2,2)$ modes enter in the quasi-linear regime for
$t < 0.5$ (see Figure~\ref{f-vachera}), and $(3,3)$, $(4,4)$, \ldots pairs are
among the first excited wave numbers. For times $t\leq 6$,
they never amount to a significant contribution to the total
poloidal magnetic energy.

While we now know which mode numbers are dominating the dynamics at 
any particular
time, it remains to identify what causes their various couplings. 
In what follows,
we sketch the physical mechanisms which trigger individual modes during the
growth and saturation phase of the 3D KH instability. It should be clear
that the {\em combined} finite volume and spectral calculations led
us to a fairly detailed insight in the intricate non-linear dynamics. 

\subsection*{Analytic results for quasi-linear regime}

First, we can make some analytic predictions for times earlier than
$t\simeq 0.5$, where the behavior is essentially `quasi-linear'. Since the
initial plasma beta is high, $\beta(t=0)\approx 120$, we can neglect magnetic
effects altogether during this phase. We employ a cylindrical coordinate system
$(R,\varphi,Z)$ centered on the jet. The total velocity field can be written
as a sum of the equilibrium shear flow, the imposed radial velocity 
perturbation, and an induced small velocity perturbation as follows:
\begin{equation}
   \vv(R,\varphi,Z;t)\simeq v_0(R)\,\hat{e}_Z+\Delta v_R(R)\, 
   \cos{\varphi}\sin\frac{2\pi Z}{L}\,\hat{e}_R+\delta\vv(R,\varphi,Z;t).
\label{q-vel}
\end{equation}
We then analyse the linear and most dominant non-linear effects in a
pure hydrodynamic description. On the jet surface $R=R_{jet}$,
$v_0(R_{jet})=0$ and 
$\Delta v_R (R_{jet}) \gg \delta v_R(R_{jet})$. In the vicinity of this
jet surface, we can safely assume that $\Delta v_R$ is constant.
We start from the hydrodynamic equations, written
in cylindrical coordinates, and employ primitive variables $\rho$, $p$, and
$\vv$. The continuity equation in cylindrical coordinates is
\begin{equation}
  \frac{\partial \rho}{\partial t}+v_R \frac{\partial \rho}{\partial R} +
  \frac{v_{\varphi}}{R}\frac{\partial \rho}{\partial \varphi}
  +v_Z \frac{\partial \rho}{\partial Z} 
  + \rho \left[\frac{1}{R}\frac{\partial (R v_R)}{\partial R}+\frac{1}{R}
  \frac{\partial v_{\varphi}}{\partial \varphi}
  +\frac{\partial v_Z}{\partial Z}\right]  =  0.
\label{q-rho1}
\end{equation}
Using expression~(\ref{q-vel}), writing 
$\rho\equiv\rho_0+\delta \rho(R,\varphi,Z,t)$, 
$p\equiv p_0+\delta p(R,\varphi,Z,t)$, and evaluating 
at $R=R_{jet}$, we get:
\begin{eqnarray}
  \frac{D \rho}{Dt}+\rho \nabla \cdot \vv = 0 & \ \ \ \Rightarrow \ \ \ 
  & \delta \rho \simeq 
  - t \frac{\rho_0}{R}\Delta v_R \cos{\varphi}\sin\frac{2\pi Z}{L} + \ldots
\label{q-rho}
\\
\delta p = c_0^2 \delta \rho & \ \ \ \Rightarrow \ \ \ 
  & \delta p \simeq
  - t \frac{\gamma p_0}{R}\Delta v_R \cos{\varphi}\sin\frac{2\pi Z}{L} + \ldots
\label{q-p} 
\end{eqnarray}
where $c_0=\sqrt{\gamma p_0/\rho_0}$ is the sound speed for the unperturbed 
configuration.

Hence, the linear pressure perturbation on the jet surface, 
in phase with the density perturbation,
follows immediately from the non-vanishing radial divergence of the imposed
perturbation 
through the term $\frac{1}{R}\frac{\partial }{\partial R}(R v_R)$.
All other terms in equation~(\ref{q-rho1}) 
lead to ${\cal O}\left(\Delta v_R \, t^2\right)$ contributions, which 
enter the dynamics later.

In the $\varphi$-component of the momentum equation, 
\begin{equation}
  \frac{\partial v_{\varphi}}{\partial t}+v_R \frac{\partial v_{\varphi}}
  {\partial R} +
  \frac{v_{\varphi}}{R}\frac{\partial v_{\varphi}}{\partial \varphi}
  +v_Z \frac{\partial v_{\varphi}}{\partial Z} 
  +\frac{v_R v_{\varphi}}{R}  =  -\frac{1}{R \rho}
  \frac{\partial p}{\partial \varphi},
\label{q-vphi1}
\end{equation}
the dominant term at $R=R_{jet}$ is the linearly induced
pressure gradient in the $\varphi$-direction, so using~(\ref{q-p}) we get:
\begin{equation}
  \frac{\partial \delta v_{\varphi}}{\partial t} = 
  - t c^2_0\frac{\Delta v_R}{R^2_{jet}}\sin{\varphi}\sin\frac{2\pi Z}{L} 
  + \ldots
\label{q-vphi}
\end{equation}
Similarly, the longitudinal velocity $v_Z$-component 
at $R=R_{jet}$ is dominated by purely linear effects, but it has two distinct
contributions. The full equation is 
\begin{equation}
  \frac{\partial v_Z}{\partial t}+v_R \frac{\partial v_Z}
  {\partial R} +
  \frac{v_{\varphi}}{R}\frac{\partial v_Z}{\partial \varphi}
  +v_Z \frac{\partial v_Z}{\partial Z} 
  = -\frac{1}{\rho}
  \frac{\partial p}{\partial Z}.
\label{q-vz1}
\end{equation}
Again, the induced
pressure gradient term along $Z$ contributes, but now also the shear
in the background flow profile has a linear term associated with it, such that
\begin{equation}
  \frac{\partial \delta v_Z}{\partial t} \simeq
  -\Delta v_R \cos{\varphi}\sin\frac{2\pi Z}{L} 
  \left.\frac{d v_0}{d\,R}\right|_{R_{jet}}
  + t c^2_0\frac{\Delta v_R}{R_{jet}}
  \frac{2\pi}{L}\cos{\varphi}\cos\frac{2\pi Z}{L} + \ldots
\label{q-vz}
\end{equation}
So far, all terms considered are purely linear. The quasi-linear effects
enter by a consideration of the equation for the radial velocity,
\begin{equation}
  \frac{\partial v_R}{\partial t}+v_R \frac{\partial v_R}
  {\partial R} +
  \frac{v_{\varphi}}{R}\frac{\partial v_R}{\partial \varphi}
  +v_Z \frac{\partial v_R}{\partial Z} 
  -\frac{v_{\varphi}^2}{R}
  = -\frac{1}{\rho}
  \frac{\partial p}{\partial R}.
\label{q-vR1}
\end{equation}
Close to the jet surface, the appropriate ordering of terms becomes
\begin{equation}
  \frac{\partial \delta v_R}{\partial t} =
  -\frac{1}{\rho_0}\frac{\partial \delta p}{\partial R}
  -\delta v_Z \Delta v_R \frac{2\pi}{L}\cos{\varphi}\cos\frac{2\pi Z}{L}
+\frac{\delta v_{\varphi}}{R_{jet}} \Delta v_R \sin\varphi \sin\frac{2\pi Z}{L}
  +\frac{(\delta v_{\varphi})^2}{R_{jet}} + \ldots
\label{q-vr1}
\end{equation}
The non-linear effects arise from the $\vv \cdot \nabla \vv$ terms. 
We can approximate~(\ref{q-vr1}), using the linear terms in the
expressions~(\ref{q-p}),~(\ref{q-vphi}) and~(\ref{q-vz}), as
\begin{eqnarray}
  \frac{\partial \delta v_R}{\partial t}  & \simeq  &
  t\,\Delta v_R \frac{c^2_0}{R^2_{jet}} \cos{\varphi}\sin\frac{2\pi Z}{L}
  +t\,(\Delta v_R)^2 \frac{d v_0}{dR}_{\mid R_{jet}} \frac{\pi}{2 L}
  \left[\sin\frac{4\pi Z}{L}+\cos(2\varphi)\sin\frac{4\pi Z}{L}\right] 
\nonumber \\
  & - &t^2 (\Delta v_R)^2 \frac{c^2_0}{2 R_{jet}}
  \left[\frac{4\pi^2}{L^2}\cos^2{\varphi}\cos^2\frac{2\pi Z}{L}
  +\frac{1}{R^2_{jet}}\sin^2{\varphi}\sin^2\frac{2\pi Z}{L}\right]
  +{\cal O}\left\{t^4 (\Delta v_R)^2\right\} 
\nonumber \\
\label{q-vr2}
\end{eqnarray}
Hence, the $(1,1)$ excitation leads to $(0,2)$, $(2,2)$ contributions in
the quasi-linear regime. The third term in expression~(\ref{q-vr2}) has
a $(2,0)$ effect in it as well, since 
$\cos^2 \alpha = (\cos(2\alpha)+1)/2$. The quasi-linear analytic reasoning
thus validates the observed ordering in time in the first-excited
linear mode numbers as seen in Fig.~\ref{f-vachera}. At the same time,
we identified the physical effects that cause their excitation.

\subsection*{Excitation of secondary KH instabilities}

Beyond this quasi-linear phase, the physical interpretation of various
non-linear coupling events must be guided by the numerical
solutions calculated with both codes. Following Fig.~\ref{f-vachera},
it remains to identify why, in addition to the
excited $(1,1)$ modes, the $m=3$ and $n=2$ wave numbers
govern the dynamics at the time of saturation at $t\approx 4$.
Moreover, the influence of the magnetic field needs to be discussed.
This is best illustrated using the picture gallery shown in Fig.~\ref{f-t4}.
 
The first frame A shows the density perturbation at $t=4$ in a horizontal
cutting plane through the jet axis. 
The sinusoidal, sideways $(1,1)$ displacement of the jet essentially
reduces to a `doubled' 2D simulation in this plane. Indeed,
at each of both intersections with the jet surface we created initial conditions
much like those used for the 2D result shown at saturation in Fig.~\ref{f-2d}.
The 3D result shows that the basic features are unaltered: 
vortical flow has redistributed mass along the
direction of the jet in periodic depletions (dark regions) and enhancements
(bright). At the center of those regions where mass 
is accumulated, a narrow lane of low-density is formed which coincides
with high magnetic fields, as the field 
gets entrained and compressed by the global
plasma circulation. There is a significant difference with the 2D result,
namely that the positioning of the resulting
saturation pattern is shifted along the jet axis. In the pure 2D simulation
from Fig.~\ref{f-2d}, 
the induced vertical pressure gradient $\frac{d \delta p}{d y}$
is in phase with 
the initial vertical velocity perturbation $\delta v_y \sim\sin{2\pi x}$.
The shear background flow introduces a force
within the shear layer which is purely along the $x$-direction. The quasi-linear
analytic treatment presented above clearly shows in equation~(\ref{q-vz})
how an extra pressure gradient term, 
out of phase with the imposed velocity perturbation,
enters the description. As a result, the observed axial shift of the pattern
ensues. 

Frame B in Fig.~\ref{f-t4} shows the same horizontal cut in a 3D
view of the simulation domain, and a specific high-density isosurface
where $\rho=1.07$. The actual range in density is now $[0.73,1.11]$.
Note how the high density zones appear as two `banana'-shaped surfaces
at diagonally opposing positions with respect to the center of the box.
The density perturbation is thus clearly dominated by a $(1,1)$ mode
number pair. The enhanced density regions coincide with zones of
excess pressure, according to equation~(\ref{q-rho}). This is best demonstrated 
by visualizing the jet surface $v_x=0$, and coloring it with its
local thermal pressure. This is done in frame C of the picture gallery.
Note the diametrically opposing high pressure zones (bright)
in the same location as the high density regions, and similarly for
the low pressure (or density) regions along the other diagonal. For the
particular aspect ratio of the simulated jet, the resulting
poloidal pressure gradient in the $\varphi$-direction on the jet surface
has induced a particular poloidal flow pattern on top and bottom of the jet
surface. This flow essentially wraps the high pressure regions around top
and bottom of the jet surface, thereby invading the low pressure zones. This
occurs along 
the positive $y$-direction in the front half of the jet, and
along the negative $y$-direction at the back half. This effectively doubles
the axial mode number $n$ of the perturbation along top and bottom of the jet.
If we make a vertical cut through the density structure containing the jet
axis, as in frame D of Fig.~\ref{f-t4}, an induced $n=2$ KH perturbation
is clearly visible. 

\subsection*{Role of the magnetic field at saturation}

So far, the role of the magnetic field is not fully explained. However,
at saturation $t=4$, the plasma beta is locally reduced to $\beta\approx
4.96$ from the initially uniform $\beta(t=0)\approx 120$, while the
Alfv\'enic Mach number $M_a$ is as low as $0.01$. Magnetic effects can no
longer be ignored. The induced flow pattern causing the excitation of a $n=2$
KH perturbation at top and bottom of the jet also entrails the field lines.
Frame E in Figure~\ref{f-t4} shows the $\log_{10}(B^2_{pol})=-1.4$ 
isosurfaces of the squared poloidal magnetic
field strength, which was zero at $t=0$. 
These are cospatial with the strongest total magnetic field zones, 
and four distinct regions are
visible: two sheet-like structures at the positions where the density is
enhanced, and two curly fibril structures along top and bottom of the jet.
The sheet structures are the immediate 3D generalization of the
low-density, high-field lanes visible in the 2D calculation shown
in Fig.~\ref{f-2d}. Note how they coincide with 
a central low-density zone within the `banana' high density
isosurface from frame B. The fibril structures are high field regions built up
by the induced flow pattern around top and bottom of the jet, discussed
earlier. They coincide with low pressure, low density regions with a similar
3D curved shape. In fact, the vertical cut of the density structure
from frame D contains three intersections of the actual 3D low density
fibril at each jet crossing. Just as in 2D, the magnetic field has become
locally dominant in the dynamics, and thereby saturates the instability
and controls the further evolution. The final frame illustrates this
clearly, where we now visualize the velocity field in a vertical cut
perpendicular to the jet axis at $x=0.5$. 
The arrows indicate the poloidal velocity
field, and a clear circulation is seen about the high field fibrils.
The contours of the $x$-component of vorticity are also showing the same
effect. The thick solid line is the $v_x=0$ jet surface, which has a
clear $m=3$ perturbation in it. Hence, the 
thermal pressure and the magnetic field cooperate to induce
$n=2$ and $m=3$ contributions that dominate the saturation behavior
along with the $(1,1)$ initial disturbance. The resulting density
variation in three space dimensions is shown in Fig.~\ref{f-rhot4}, and
can now be interpreted from the preceding exposition.

\subsection{Case $m=2$}

Figure~\ref{f-vachera2} shows the development and saturation behavior
of the KH unstable jet, when the initial excitation has a $\cos{2\varphi}
\sin{2\pi Z/L}$ variation. 
Figure~\ref{f-rhot4m2} gives a 3D impression of
the density variation at $t=4$ for this case, and there is significant
fine structure in all spatial directions. 

The information on the excitation and coupling events between
linear modes can again be extracted directly from the HERA calculation, 
and used for the physical interpretation of the non-linear dynamics. A
quasi-linear description, 
completely analogous to the one for $m=1$, predicts the
initial excitation of the $(0,2)$, $(4,2)$, and $(4,0)$ modes. The later
evolution and saturation is dominated by the imposed $(2,1)$ and the
excited $(4,0)$ mode pair. The discussion of the simpler $m=1$ case in terms
of pressure-gradient induced effects and the role of the magnetic field
can be carried over to the present configuration. Note that the initial
$(2,1)$ perturbation leads to four distinct zones where flows converge,
building up high density, high pressure regions. Instead of the two diagonally
opposing regions as shown in frame B of Figure~\ref{f-t4}, we now have
two regions in the front half of the jet $0<x<0.5$, and two in the back
half $0.5<x<1$. The front two are at the top and the bottom of the
jet, with the back ones at the left and right edge. Figure~\ref{f-rhot4m2}
shows three of the enhanced density zones (bright) using various cutting
planes through the simulation box at $t=4$. As a result of the
associated pressure variations about the jet surface, flows are set up
which again entrail magnetic field lines. As a direct generalization
of frame E in Figure~\ref{f-t4}, one ends up with sheets of high
magnetic fields intersecting the four compression zones, and a set of
four fibril structures curled in between these zones as a secondary effect
from the pressure induced flows. The effects on the density structure
are well represented in Figure~\ref{f-rhot4m2}. Values range from
$0.77$ to $1.10\,$.

\subsection{Further evolution}\label{ss-evol}

The evolution beyond the non-linear saturation is governed by many more
mode numbers, as clearly seen in Figures~\ref{f-vachera}-\ref{f-vachera2}.
The regions where the magnetic field dominates essentially break up the
jet, as identified by its $v_x=0$ isosurface. A transition to more
turbulent flow sets in, but quantitative statements about this transition
would require higher resolution 3D studies. These must assess the role
played by numerical dissipation, not only in the way it affects the 
momentum evolution (viscous effects), but also in the developing
fine-scale magnetic field (through resistive studies). A combined approach,
like the simultaneous finite volume and spectral 
calculations presented here, will certainly
help in sorting out physics from possible numerical artefacts.

\section{Conclusions}\label{s-concl}

Both the $m=1$ and the $m=2$ case study concentrated on the
growth and saturation phase of KH unstable sheared magnetized jets.
For the chosen parameters, a detailed account of the non-linear
excitation of several $(m,n)$ 2D Fourier modes was presented. 
The combined finite volume
and spectral calculations allowed for an in-depth analysis of the physical
mechanisms behind non-linear mode couplings. The initially weak magnetic field
could be ignored to make analytic predictions about the initial excitation
of several mode pairs during the quasi-linear phase. The field eventually
becomes locally dominant, and controls the 3D density structure at saturation.
At the particular aspect ratio $L/R_{jet}=2$, we saw how an $(m,n)=(1,1)$
kink perturbation of the jet leads to a secondary KH instability, characterized
by an axial $n=2$ wave number, along top and bottom of the jet. 
The induced poloidal pressure gradient triggers this instability. The magnetic
field gets compressed in narrow sheets that intersect
$(1,1)$ compression zones, and in 3D fibrils due to these secondary KH flows.
These, in turn, cause an $m=3$ deformation of the jet surface.
Similar effects were present in the $(m,n)=(2,1)$ case study.

Clearly, the initial plasma beta, the strength of the shear flow, and the
aspect ratio of the jet all play an important role in this process.
A detailed parameter study should explore their influence, 
along the lines of previous 2D studies~\cite{vac-kh2d}. 
The particular choice of parameter values used here (subsonic, high beta)
could occur in the lower atmospheric regions of stars like our Sun: 
the high beta photospheric and chromospheric layers possess a 
variety of magnetically modified, shear flow regimes. 
We can safely speculate that magnetic effects will be important in
highly supersonic, extragalactic jets as well. Initially weak fields can become
locally dominant, e.g. through shock compression. In turn, the field may then
dictate the type of shock interactions possible to the flow.
Future investigations
can look at specific geometries (flaring flux tubes, solar coronal loops) and
adjust field strengths, shear strengths, include gravitation, etc.
Instead of a uniform, weak $\BB$ field parallel to the jet, the
initial magnetic field configuration could contain current sheet(s). It is
then possible to investigate
tearing unstable situations in conjunction with background
shear flows. Such effects have recently been studied
in 3D incompressible MHD calculations
of a magnetized wake~\cite{dahl3d} for studying the slow component 
of the solar wind. The effects of compressibility, which we analyzed for
cases whithout initial current sheets, should be quantified in
subsequent work.

\acknowledgements

This work was performed as part of the research program of the association
agreement of Euratom and the `Stichting voor Fundamenteel Onderzoek
der Materie' (FOM) with financial support from  the `Nederlandse
Organisatie voor Wetenschappelijk Onderzoek' (NWO) and Euratom.
It was part of the project on `Parallel Computational
Magneto-Fluid Dynamics', funded by the NWO Priority Program on
Massive Parallel Computing and coordinated by Prof.\ J.P.~Goedbloed.
It was sponsered by the `Stichting Nationale Computerfaciliteiten'
(NCF) for the use of supercomputer facilities (Cray C90, Cray T3E, IBM SP).
GT currently receives a postdoctoral fellowship (D 25519) from the
Hungarian Science Foundation (OTKA) and a Bolyai fellowship from the
Hungarian Academy of Sciences. He also thanks FOM for support
during his visit.

\newpage
{\bf Figure Captions:}

\vspace*{0.5cm}
{\bf Figure 1:}
The density structure after four transverse sound travel times,
in a magnetically modified 2D Kelvin-Helmholtz evolution. 
At this time, the instability is non-linearly saturated.
Dark lanes of low density where the magnetic field is intensified
intersect high density (bright) regions, as seen at the periodic edges.

\vspace*{0.5cm}
{\bf Figure 2:}
The time history of the scaled
poloidal kinetic  $E^{pol}_{kin}(t)$ ($\times 10$, thick dashed) and magnetic 
$E^{pol}_{mag}(t)$ (thick solid) energy, 
for the $m=1=n$ case,
from both the VAC 
and the HERA calculations. The growth and the non-linear 
saturation as identified
by the first maximum in $E^{pol}_{kin}(t)$ agree closely. The contributions
from the most important 
individual mode number pairs $(m,n)$ to $E^{pol}_{mag}(t)$ are
plotted as well, and the linestyle varies with the azimuthal mode number $m$:
thin solid for $m=0$ (or $m\geq 6$), dotted $m=1$, thin dashed $m=2$,
dash-dotted $m=3$, $-\cdot\cdot\cdot -$ for $m=4$, long dashes $m=5$. 

\begin{figure}
\begin{center}
\FIG{
\resizebox{\textwidth}{!}
{\includegraphics{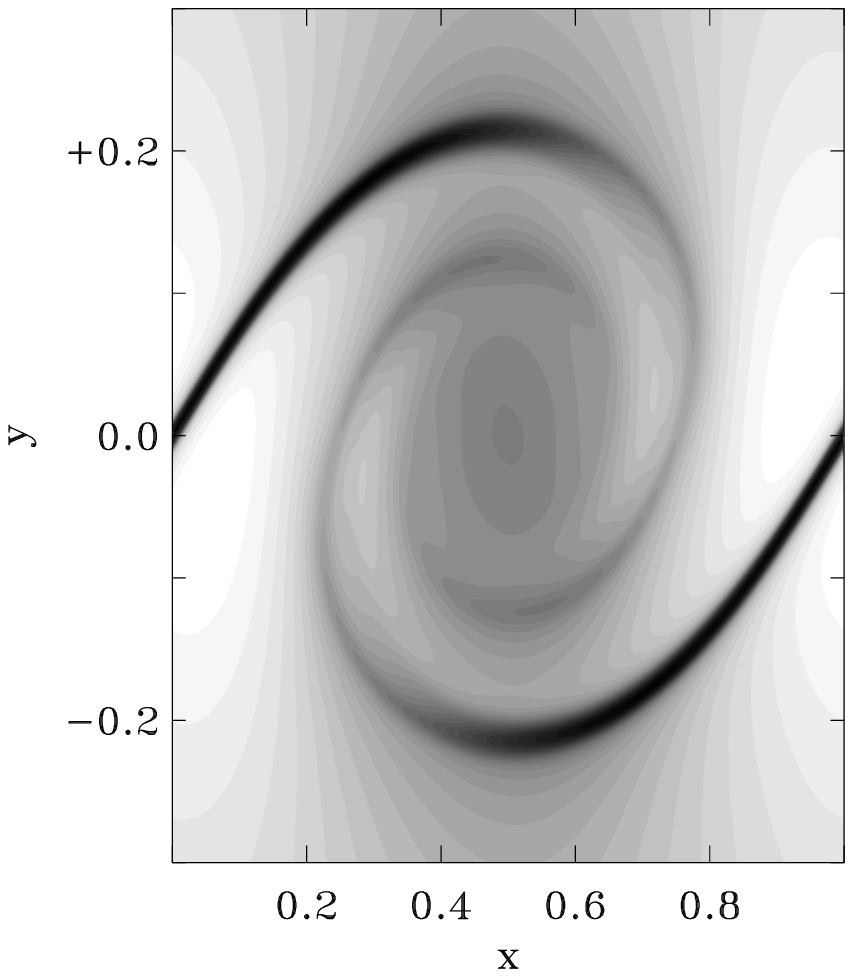}}
}
\end{center}
\caption{}
\label{f-2d}
\end{figure}

\begin{figure}
\begin{center}
\FIG{
\resizebox{!}{0.8\textheight}
{\includegraphics{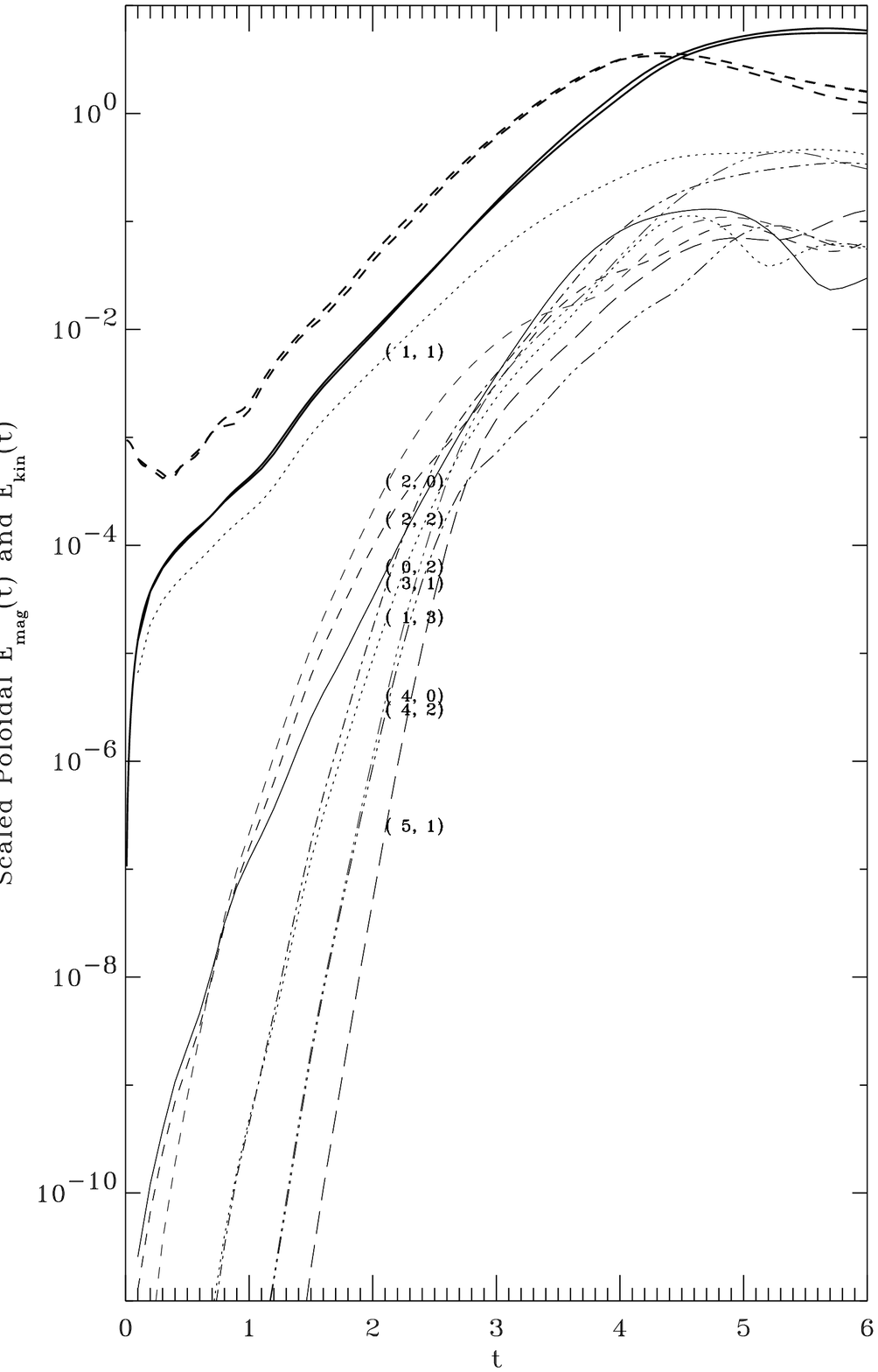}}
}
\end{center}
\caption{}
\label{f-vachera}
\end{figure}

\begin{figure}
\begin{center}
\FIG{
\resizebox{\textwidth}{!}
{\includegraphics{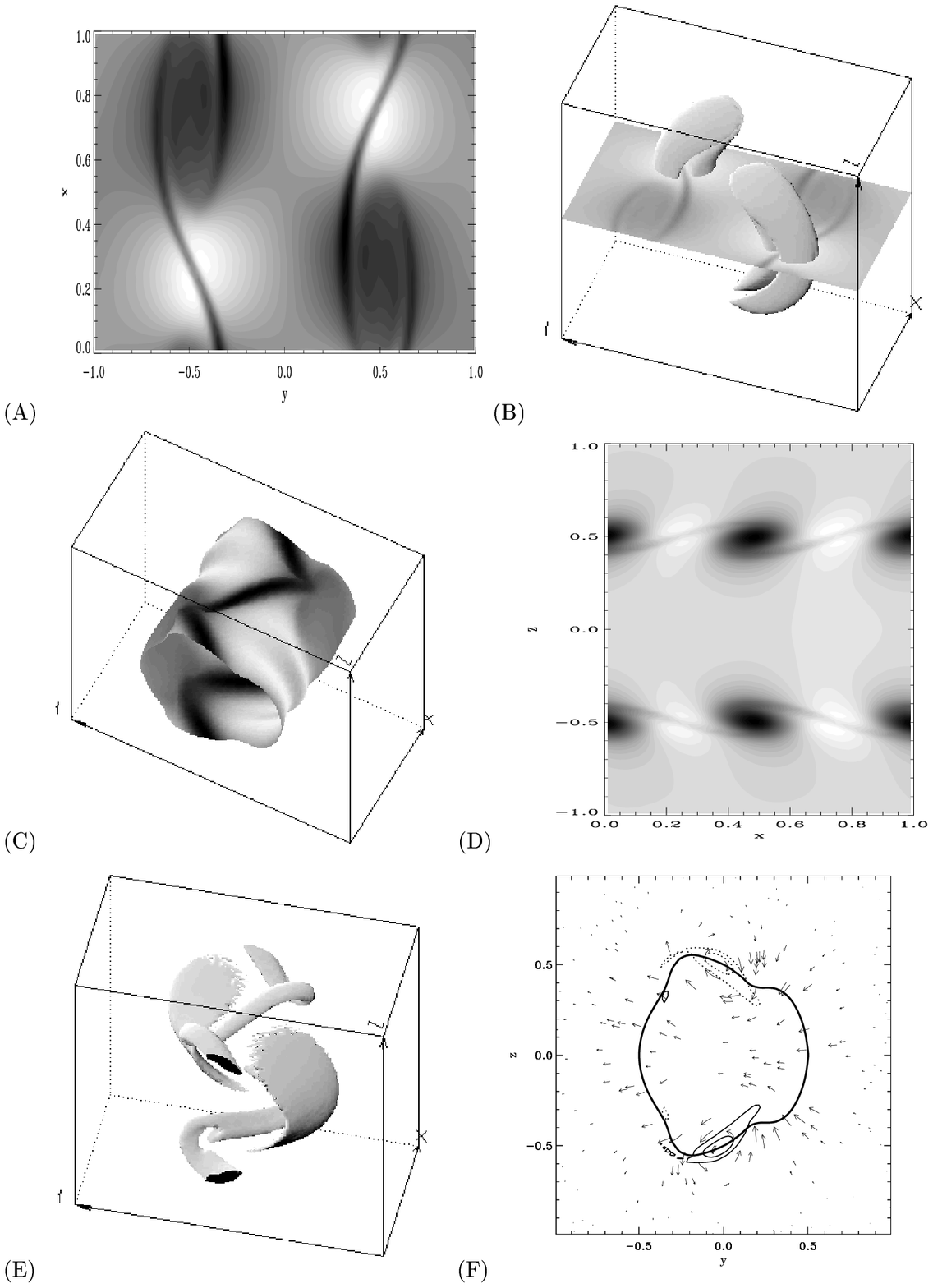}}
}
\end{center}
\caption{}
\label{f-t4}
\end{figure}

\begin{figure}
\begin{center}
\FIG{
\resizebox{\textwidth}{!}
{\includegraphics{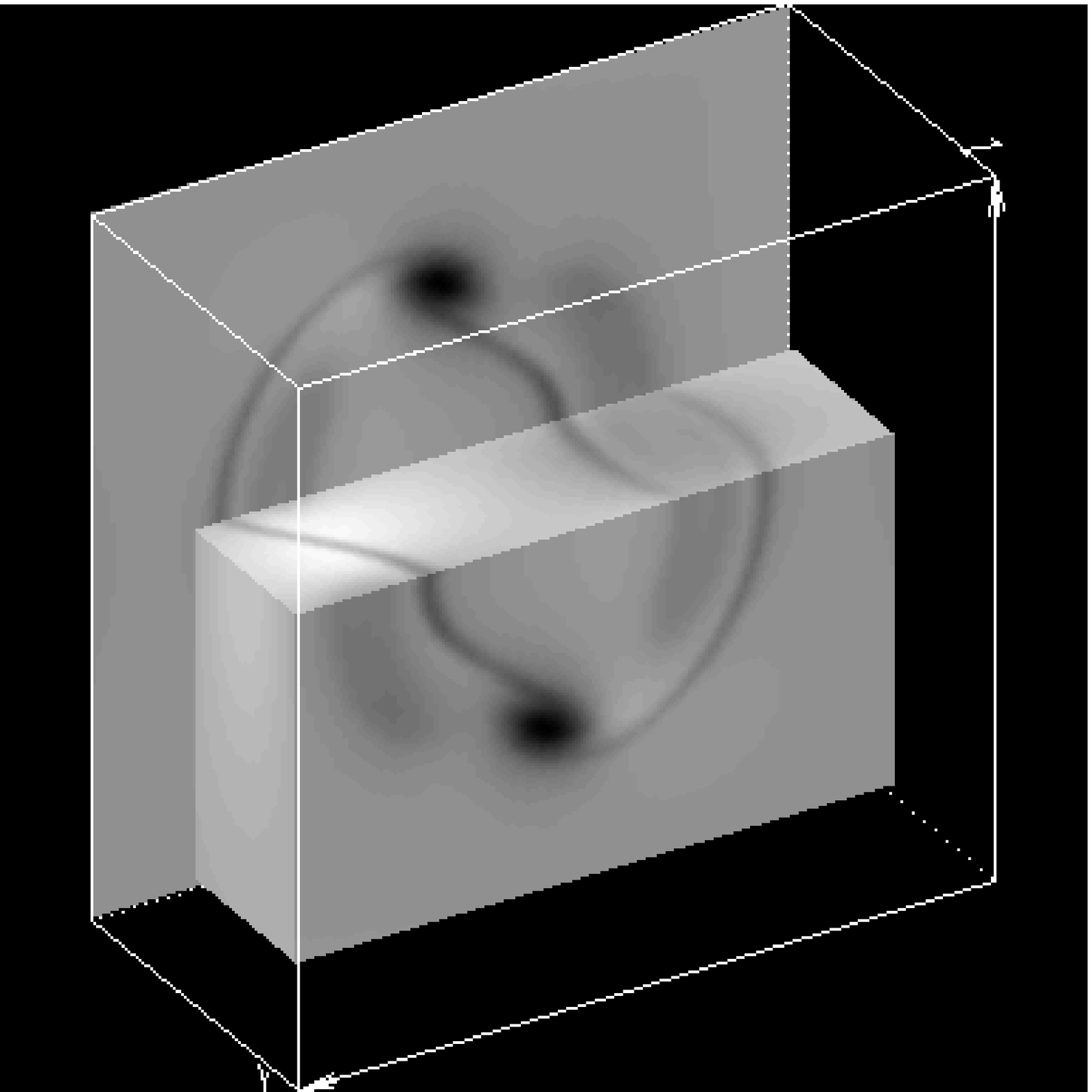}}
}
\end{center}
\caption{}
\label{f-rhot4}
\end{figure}

\begin{figure}
\begin{center}
\FIG{
\resizebox{!}{0.8\textheight}
{\includegraphics{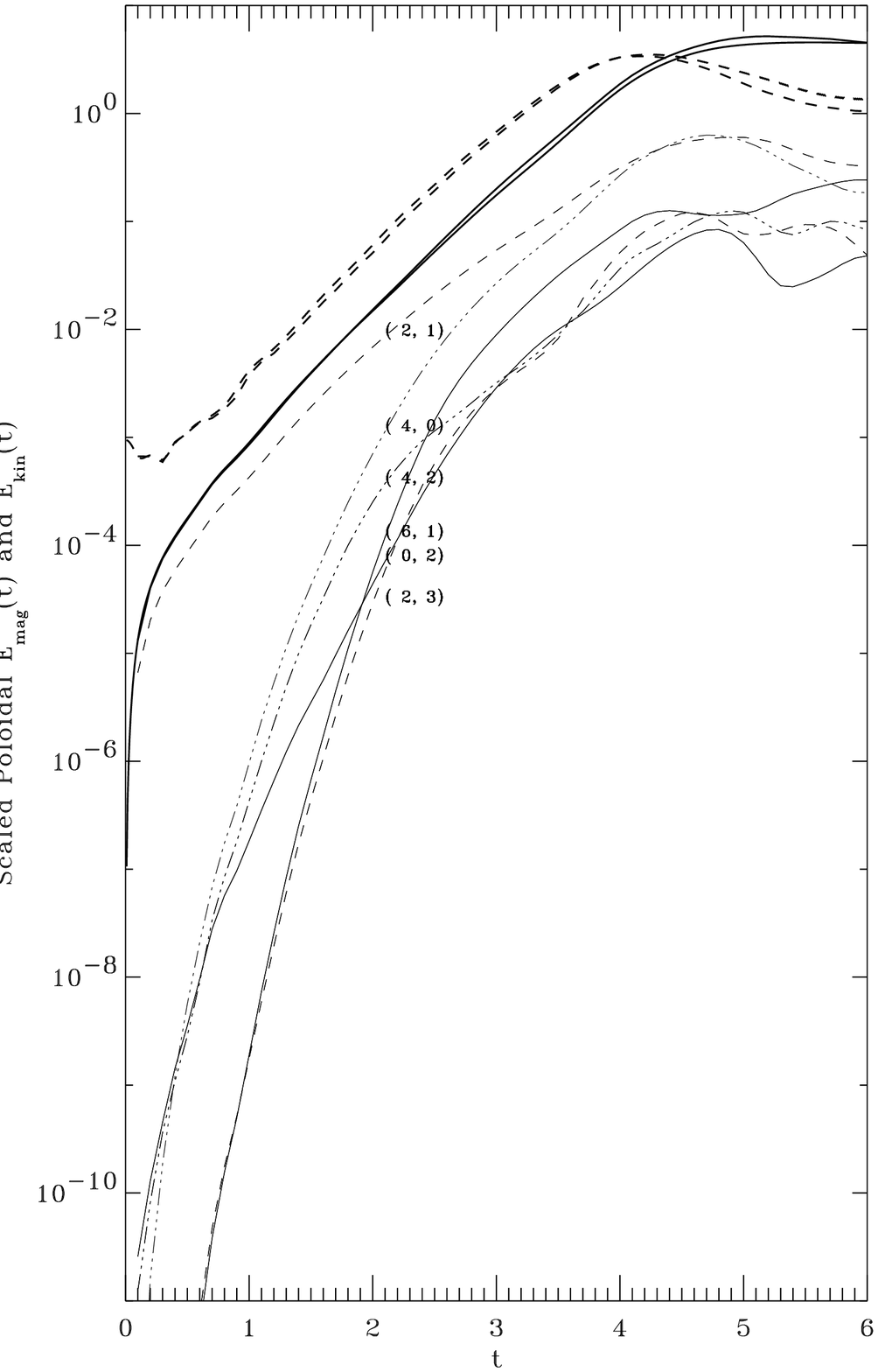}}
}
\end{center}
\caption{}
\label{f-vachera2}
\end{figure}

\begin{figure}
\begin{center}
\FIG{
\resizebox{\textwidth}{!}
{\includegraphics{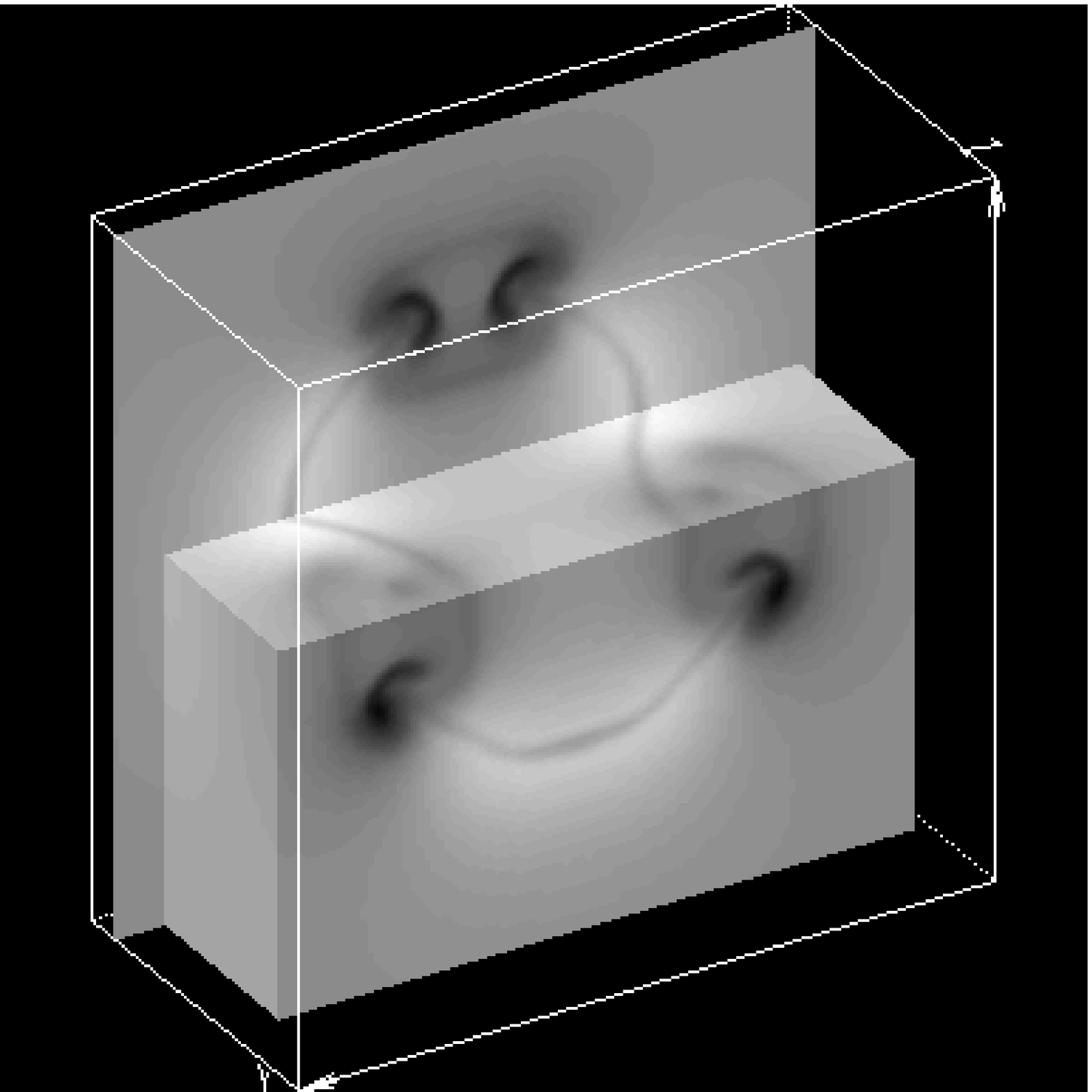}}
}
\end{center}
\caption{}
\label{f-rhot4m2}
\end{figure}

\vspace*{0.5cm}
{\bf Figure 3:}
Picture gallery for the $m=1$ case at saturation $t=4$. Frame
(A): the density structure in a horizontal cut through the jet axis, to be
compared to the 2D case in Figure~\ref{f-2d}.
(B): the same cut as in frame (A), with the $\rho=1.07$ high density
isosurface. Frame (C): the deformed jet surface $v_x=0$, colored with thermal
pressure. Frame (D): the density structure in a vertical cut containing
the jet axis, demonstrating the induced $n=2$ KH instability at top and bottom
of the jet. (E): $\log_{10}(B^{2}_{pol})=-1.4$ 
isosurfaces, showing the locations
of magnetic field dominated dynamics. Frame (F): a vertical cut 
perpendicular to the jet axis, showing the poloidal flow field as
vectors, the $v_x=0$ jet surface as a thick solid line, and contours of
$(\nabla \times \vv)_x$.

\vspace*{0.5cm}
{\bf Figure 4:}
A rotated view of the 3D density structure for the $m=1$ case 
at saturation $t=4$. The dark regions are sites of
low density, resulting from dynamically strong magnetic fields.

\vspace*{0.5cm}
{\bf Figure 5:}
As in Figure~2, for the $m=2$ case.

\vspace*{0.5cm}
{\bf Figure 6:}
The $m=2$ equivalent of Figure~4.

\newpage

\begin{references}
\bibitem{chandra}
  S. Chandrasekhar, 
  {\em Hydrodynamic and Hydromagnetic Stability,}
  Oxford University Press, New York (1961).
\bibitem{blumen}
  W. Blumen, 
  \PREP{Shear layer instability of an inviscid compressible fluid.}
  J. Fluid Mech. {\bf 40}, 769 (1970).
\bibitem{vac-kh2d}
   R. Keppens, G. T\'oth, R.H.J. Westermann, and J.P. Goedbloed,
   Growth and saturation of the Kelvin-Helmholtz instability with
   parallel and anti-parallel magnetic fields,
   accepted by J. Plasma Phys. (1998).
\bibitem{frank}
 A. Frank, T.W. Jones, D. Ryu, and J.B. Gaalaas, 
 \PREP{The magnetohydrodynamic Kelvin-Helmholtz instability: A
  two-dimensional study.}
 Astrophys. J. {\bf 460}, 777\PREP{-793} (1996).
\bibitem{jones}
 T.W. Jones, J.B. Gaalaas, D. Ryu, and A. Frank,
 \PREP{The MHD Kelvin-Helmholtz Instability. II. The Roles of Weak and
 oblique Fields in Planar Flows.}
 Astrophys. J. {\bf 482}, 230 (1997).
\bibitem{dahl97}
 R.B. Dahlburg, P. Boncinelli, and G. Einaudi, 
 \PREP{The evolution of plane current-vortex sheets.}
 Phys. Plasmas {\bf 4}, 1213\PREP{-1226} (1997).
\bibitem{dahl98}
 R.B. Dahlburg, 
 \PREP{On the nonlinear mechanics of magnetohydrodynamic stability.}
 Phys. Plasmas {\bf 5}, 133\PREP{-139} (1998).
\bibitem{malag}
 A. Malagoli, G. Bodo, and R. Rosner, 
   \PREP{On the nonlinear evolution of magnetohydrodynamic Kelvin-Helmholtz
   instabilities.}
   Astrophys. J. {\bf 456}, 708\PREP{-716} (1996).
\bibitem{miura}
 A. Miura, 
  Phys. Plasmas {\bf 4}, 2871\PREP{-2885} (1997).
\bibitem{miupr}
 A. Miura and P.L. Pritchett, 
   \PREP{
   Nonlocal stability analysis of the MHD Kelvin-Helmholtz instability
   in a Compressible plasma.}
   J. Geophys. Res. {\bf 87}, 7431\PREP{-7444} (1982).
\bibitem{min}
 K.W. Min,
 \PREP{Simulation of the Kelvin-Helmholtz Instability in the Magnetized 
  Slab Jet.}
 Astrophys. J. {\bf 482}, 733 (1997).
\bibitem{bodo}
 G. Bodo, P. Rossi, S. Massaglia, A. Ferrari, A. Malagoli, and R. Rosner, 
 Astron. \& Astrophys. {\bf 333}, 1117 (1998).
\bibitem{frank2}
 A. Frank, D. Ryu, T.W. Jones, A. Noriega-Crespo,
 \PREP{Effects of Cooling on the Propagation of Magnetized Jets.}
 Astrophys. J. Lett. {\bf 494}, L79 (1998).
\bibitem{nishi1}
 K.I. Nishikawa, S. Koide, J.I. Sakai, D.M. Christodoulou,
   H. Sol, and R.L. Mutel, 
   Astrophys. J. Lett. {\bf 483}, L45 (1998).
\bibitem{nishikawa}
   K.I. Nishikawa, S. Koide, J.I. Sakai, D.M. Christodoulou,
   H. Sol, and R.L. Mutel, 
   Astrophys. J. {\bf 498}, 166 (1998).
\bibitem{hardee}
  P.E. Hardee, D.A. Clarke, A. Rosen,
 \PREP{Dynamics and Structure of Three-dimensional Poloidally Magnetized
  Supermagnetosonic Jets.}
 Astrophys. J. {\bf 485}, 533 (1997).
\bibitem{tothcrete}
   G.~T\'oth,
   \PREP{A general code for modeling MHD flows on parallel computers:
   Versatile Advection Code,}
   Astrophys. Lett. \& Comm. {\bf 34}, 245 (1996).
\bibitem{tothvienna}
   G. T\'oth,
   \PREP{Versatile Advection Code,}
   in {\it High Performance Computing and Networking,
   Proceedings HPCN Europe 1997, 
   Lecture Notes in Computer Science,} Vol. 1225, 
   edited by B.~Hertzberger and P.~Sloot, 
   (Springer-Verlag, Berlin, 1997), pp. 253-262.
\bibitem{hera-hpcn97}
   R.~Keppens, S. Poedts, P.M.~Meijer, and J.P.~Goedbloed,
   \PREP{A Data Parallel Pseudo-Spectral Semi-Implicit 
         Magnetohydrodynamics Code,}
   in {\it High Performance Computing and Networking,
   Proceedings HPCN Europe 1997, 
   Lecture Notes in Computer Science,} Vol. 1225, 
   edited by B.~Hertzberger and P.~Sloot, 
   (Springer-Verlag, Berlin, 1997), pp. 190-199.
\bibitem{hera-hpcn98}
   R.~Keppens, S.~Poedts, and J.P.~Goedbloed,
  \PREP{Data parallel simulations of the magnetohydrodynamics of plasma loops,}
   in {\it High Performance Computing and Networking,
   Proceedings HPCN Europe 1998, 
   Lecture Notes in Computer Science,} Vol. 1401, 
   edited by P. Sloot, M. Bubak, and B. Hertzberger, 
   (Springer-Verlag, Berlin, 1998), pp. 233-241.
\bibitem{tothodstrcil}
   G. T\'oth and D. Odstr\v cil,
   \PREP{Comparison of some Flux Corrected Transport
   and Total Variation Diminishing Numerical Schemes
   for Hydrodynamic and Magnetohydrodynamic Problems,}
   J. Comput. Phys. {\bf 128}, 82 (1996).
\bibitem{implvacA}
   R.~Keppens, G.~T\'oth, M.A.~Botchev, and A. van der Ploeg,
   Implicit and Semi-Implicit Schemes: algorithms,
   submitted for publication to Int. J. for Numer. Meth. in Fluids (1998).
\bibitem{implvacB}
   G. T\'oth, R. Keppens, and M.A. Botchev,
   Astron. \& Astrophys. {\bf 332}, 1159 (1998).
\bibitem{wind98}
   R. Keppens and J.P. Goedbloed, 
   Numerical simulations of stellar winds: polytropic models,
   Astron. \& Astrophys. {\bf 342}, to appear (1999).
\bibitem{porto98}
   R. Keppens and G. T\'{o}th,
   \PREP{Simulating Magnetized Plasmas with the Versatile Advection Code,}
   in {\it Proceedings of VECPAR'98 (3rd international meeting on VECtor
   and PARallel processing),} Porto, Portugal, 1998, {\it Lecture
   Notes in Computer Science} (Springer-Verlag, Berlin, 1999) to appear.
\bibitem{poedts}
   S. Poedts, G. T\'oth, A.J.C.~Beli\"en, and J.P. Goedbloed,
   \PREP{Nonlinear MHD simulations of wave dissipation in flux tubes,}
   Solar Phys. {\bf 172}, 45 (1997).
\bibitem{hpfart}
   R. Keppens and G. T\'oth, 
   Using High Performance Fortran for Magnetohydrodynamic simulations,
   submitted to Parallel Computing (1998).
\bibitem{vac-hpcn98}
   G.~T\'oth and R. Keppens,
   \PREP{Comparison of Different Computer Platforms for Running
   the Versatile Advection Code,}
   in {\it High Performance Computing and Networking,
   Proceedings HPCN Europe 1998, 
   Lecture Notes in Computer Science,} Vol. 1401, 
   edited by P. Sloot, M. Bubak, and B. Hertzberger, 
   (Springer-Verlag, Berlin, 1998), pp. 368-376.
\bibitem{harten}
   A. Harten,
   \PREP{High Resolution Schemes for Hyperbolic Conservation Laws,}
   J. Comput. Phys. {\bf 49}, 357 (1983).
\bibitem{collwood}
   P. Collela and P.R. Woodward,
   \PREP{The Piecewise Parabolic Method (PPM) for Gas-Dynamical simulations,}
   J. Comput. Phys. {\bf 54}, 174 (1984).
\bibitem{roe}
   P.L. Roe,
   \PREP{Approximate Riemann Solvers, Parameter Vectors, 
  and Difference Schemes,}
   J. Comput. Phys. {\bf 43}, 357 (1981).
\bibitem{brackbarn}
   J.U.~Brackbill and D.C.~Barnes,
   \PREP{Note: The Effect of Nonzero $\nabla\cdot\BB$ on the Numerical Solution
   of the Magnetohydrodynamic Equations,}
   J. Comput. Phys. {\bf 35}, 426 (1980).
\bibitem{dahl3d}
 R.B. Dahlburg, J.T. Karpen, G. Einaudi, and P. Boncinelli,
 \PREP{Acceleration of the slow solar wind}, in
 {\it Solar Jets and Coronal Plumes, Proceedings of International Meeting at
Guadeloupe,} DOM, France, 23-26 February 98, ESA Publications Division,
ESTEC, Noordwijk, The Netherlands,
ESA-SP-421 (May 1998), pp. 199-205.
\end{references}
\end{document}